\documentclass[aps,prl,twocolumn,superscriptaddress,altaffillsymbol]{revtex4-1}

\usepackage[autostyle=true]{csquotes}

\usepackage{amsmath}
\usepackage{amsfonts}
\usepackage{mathtools}

\usepackage[normalem]{ulem}
\usepackage{color}
\usepackage{graphicx}
\usepackage{dcolumn}
\usepackage{bm}
\usepackage{mathtools}
\usepackage{siunitx}

\usepackage[unicode=true,
bookmarks=true,bookmarksnumbered=true,bookmarksopen=true,bookmarksopenlevel=2,
breaklinks=false,pdfborder={0 0 1},backref=false,colorlinks=true,allcolors=blue]{hyperref}

\begin{document}

\author{Minghao Li}
\affiliation{Universit\'e de Strasbourg, CNRS, Institut de Science et d'Ing\'enierie Supramol\'eculaires, UMR 7006, 67000 Strasbourg, France.}
\author{Oussama Sentissi}
\affiliation{Universit\'e de Strasbourg, CNRS, Institut de Science et d'Ing\'enierie Supramol\'eculaires, UMR 7006, 67000 Strasbourg, France.}
\author{Stefano Azzini}
\altaffiliation[Present address: ]{Dipartimento di Fisica, Trento University, I-38123, Povo, Trento, Italy.}
\affiliation{Universit\'e de Strasbourg, CNRS, Institut de Science et d'Ing\'enierie Supramol\'eculaires, UMR 7006, 67000 Strasbourg, France.}
\author{Gabriel Schnoering}
\altaffiliation[Present address: ]{Laboratory of Thermodynamics in Emerging Technologies, Sonneggstrasse 3, 8092 Zurich, Switzerland.}
\affiliation{Universit\'e de Strasbourg, CNRS, Institut de Science et d'Ing\'enierie Supramol\'eculaires, UMR 7006, 67000 Strasbourg, France.}
\author{Antoine Canaguier-Durand}
\altaffiliation[Present address: ]{Saint-Gobain Research Paris, 39 quai Lucien Lefranc, 93303 Aubervilliers, France.}
\affiliation{Laboratoire Kastler-Brossel, Sorbonne Universit\'e, CNRS, ENS-PSL University, Coll\`ege de France, Paris, France.}
\author{Cyriaque Genet}
\email[]{genet@unistra.fr}
\affiliation{Universit\'e de Strasbourg, CNRS, Institut de Science et d'Ing\'enierie Supramol\'eculaires, UMR 7006, 67000 Strasbourg, France.}

\title{Subfemtonewton force fields measured with ergodic Brownian ensembles}

\begin{abstract}
We demonstrate that radiation pressure force fields can be measured and reconstructed with a resolution of $0.3$ fN (at a $99.7\%$ confidence level) using an ergodic ensemble of overdamped colloidal particles. The outstanding force resolution level is provided by the large size of the statistical ensemble built by recording all displacements from all diffusing particles, regardless of trajectory and time. This is only possible because the noise driving the particles is thermal, white and stationary, so that the colloidal system is ergodic, as we carefully verify. Using an ergodic colloidal dispersion for performing ultra-sensitive measurements of external forces is not limited to non-conservative optical force fields. Our experiments therefore give way to interesting opportunities in the context of weak force measurements in fluids.
\end{abstract}

\maketitle 

\section{Introduction}

That light can exert a pressure force on an illuminated object is one central prediction of Maxwell's theory \cite{Maxwell1873} that immediately challenged a few experimentalists (see \cite{Worrall1982} and references therein). But their initial attempts in measuring radiation pressure were mostly hindered by thermal effects induced by the illuminating light, such as convective and radiometric forces. The first positive experimental demonstration of light radiation pressure was achieved by P.N. Lebedev in 1899, with results published in a 1901 article \cite{Lebedev1901}. In this tour-de-force experiment performed with reflecting winglets suspended on torsional balances, Lebedev was able to measure radiation pressure forces down to $3\times 10^{-10}$ N, with an accuracy better than $6\%$.

Today, the interest in measuring radiation pressure remains. The non-conservative character of radiation pression indeed plays a central role in the generation and control of complex optical force fields \cite{He1995,Cuche2012,Kall2014,Ma2015,CanaguierPRA2013,CapassoPRL2016,SchnoeringPRAppl2019}, with non-trivial optomechanical effects recently discussed \cite{Roichman2008,Florin2009,Brzobohaty2013,Gloppe2014,CanaguierPRA2015,Dogariu2015,Dogariu2017,Brasselet2018,MangeatPRE2019}. Usually, experiments measuring external force fields involve Brownian probes and monitor, through different means and methods, the shifts of the statistical distributions of displacement of the probes from their stable equilibrium positions induced by the force fields.

In this paper, we show that an \textit{ensemble} of colloidal particles diffusing in the overdamped regime within an external non-conservative force field can be advantageously used for ultra-sensitive radiation pressure measurements. We demonstrate indeed an unprecedented level of force resolution of $0.3$ fN (within a thermally limited $99.7\%$ confidence interval) by exploiting the large size of the statistical ensemble of Brownian displacements supplied by the colloidal dispersion. We emphasize that such large statistical ensembles gather displacements recorded on different trajectories and at different times, so that our force measurement method crucially relies on the ergodic hypothesis. In order to qualify our colloidal dispersion as an ergodic system, and thus validate our strategy, we demonstrate the thermal and stationary nature of the noise driving our Brownian colloidal ensemble. 

One cannot fail to highlight the importance of the ergodic hypothesis which is needed for averaging out the contribution of the thermal when determining the force, as we explain below. The hypothesis is pivotal in many recent experiments, in particular those involving optical traps in fluids, but it is scarcely verified. In this work, we close this loophole by resorting to an appropriate measure of ergodicity. This measure yields a precise confirmation of the ergodic hypothesis for our experiments, all systematic and tracking errors being accounted for. 

\section{Experimental setup}

\begin{figure}[htb]
  \centering{
    \includegraphics[width=0.75\linewidth]{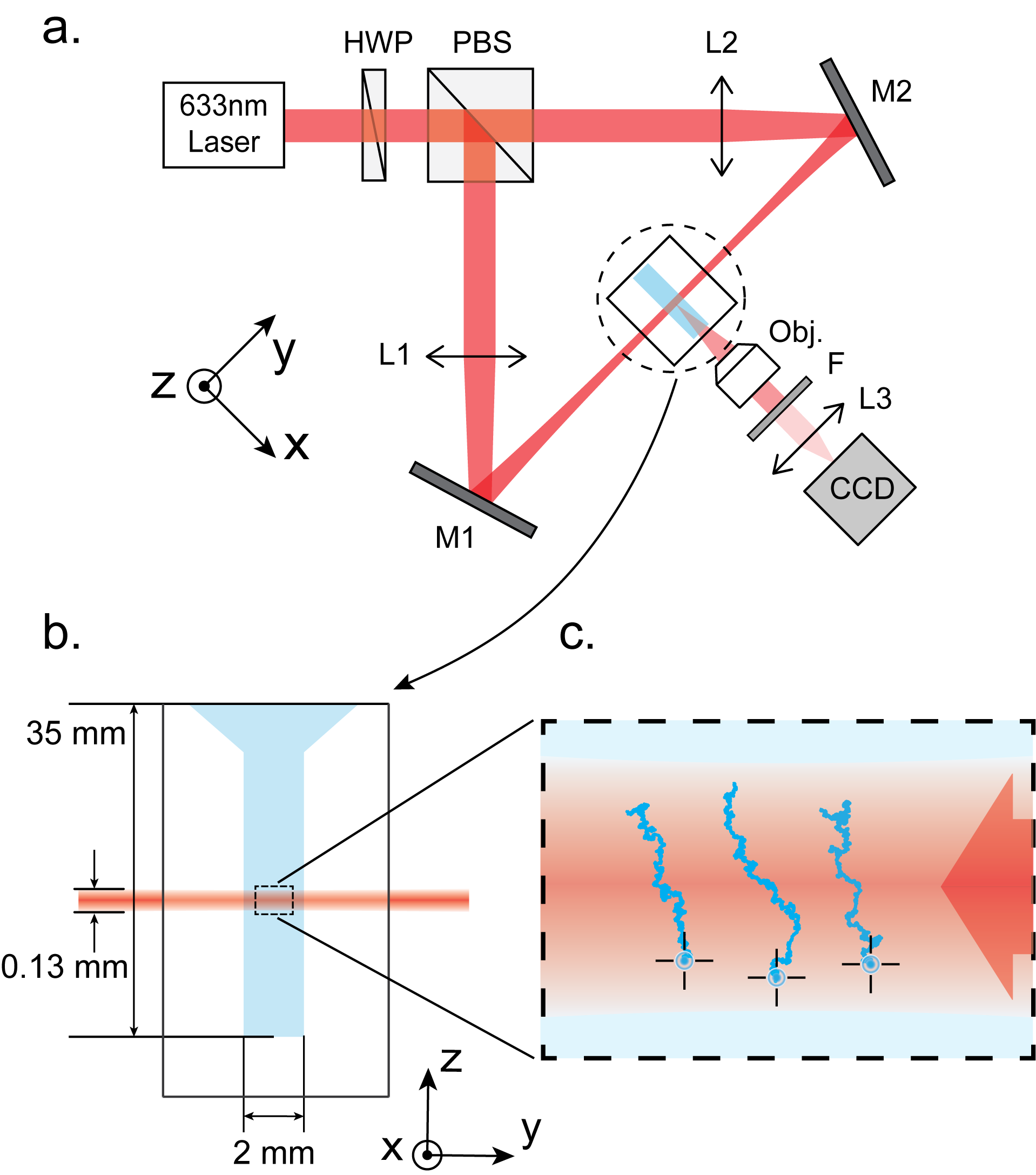}}
  \caption{Schematics of the experimental setup.  (a) A linearly polarized single-mode TEM$_{00}$ horizontal laser beam (633 nm, 200 mW) can, using a half-wave plate (HWP), a polarizing beam-splitter (PBS) and two mirrors (M1,M2), either be directed towards the sample along the $y$ axis (single-beam mode) or be splitted in two non-interfering (crossed polarizations) counter-propagating beams of identical intensity (using the HWP for the fine balance). A microscope objective (NA=0.25, $20\times$) collects the fluorescence of dye-doped melamine spheres (diameter $d=940 \pm 50$ nm, from Micro-particles GmbH) diffusing in water inside the cell. The particles are imaged on a CCD camera at a frame rate $f=120$ Hz and tracked using an algorithm adapted from \cite{Algo}. A filter F eliminates any stray light coming from the laser. (b) The sample consists of a quartz cuvette of dimensions --$10(x)\times 2(y)\times 35(z)$ mm$^3$-- chosen such that the imaging region-of-interest is located far away from any wall, allowing us to neglect safely any perturbation of the wall on the diffusion dynamics. The profile of the laser sent through the cuvette is set using $400$ mm focusing lenses (L1,L2) for large waist $w_0=65~\mu$m and Rayleigh range $z_r=18$ mm so that the illumination is a plane-wave with a Gaussian distribution of intensity in the transverse $z$ direction. (c) Region-of-interest --350$(y)\times$250$(z)\mu$m$^2$-- imaged (using lens L3) from the center of the illumination zone. The right hand side arrow corresponds to the laser propagation direction. In this single-beam mode using a fairly high intensity $0.37$ kW$\cdot{\rm cm}^{-2}$, the recorded trajectories clearly reveal the shift induced along $y$ by radiation pressure. In such conditions of high intensity, the diffusion along $z$ results from the combination of sedimentation and laser-induced convection --see main text and Appendix A. The crosses indicate the initial positions taken by the particles: due to strong convective drag, the particles are diffusing upwards.
 }
  \label{fig1}
\end{figure}

Our experiments consist in illuminating with a horizontal laser beam a colloidal dispersion of micron-sized melamine spheres, diffusing and sedimenting inside a cuvette filled with water. Coming from one side of the cuvette, the laser beam induces a radiation pressure on the particles that modifies their diffusion dynamics. As discussed in details below, this mechanical action can be analyzed by looking at colloidal trajectories in real-time, recorded by tracking the successive positions of the particles, which are fluorescent (weakly dye-doped). The large number of particles leads to collect a large number of trajectories, and thereby to provide a large statistical ensemble of single-step displacements on which a high-precision motional analysis can be performed. To do so, the optical setup, described in details in Fig. \ref{fig1}, has important features. 

First, the cuvette has large dimensions compared to the size of the imaging region-of-interest that only extends over a small central region far away from all walls. This, together with the small-volume fraction of the colloidal dispersion allows us to neglect the influence of possible boundary-wall and particle-interaction effects on the colloidal diffusion dynamics. Importantly, such dimensions also insure that sedimentation and possible laser-induced convection effects (see Appendix A) are laminar, only acting on the vertical $z$ axis. These effects thereby are perfectly projected out by looking at Brownian motion along the $y$ axis only. This is a central point in our methodology, as discussed below.

Second, the illumination conditions are set so that, within the imaging region-of-interest, the Gaussian profile of the laser beam is uniform along the horizontal optical $y$ axis (i.e. a Rayleigh length much larger than the width of the cuvette) with a waist much larger than the diameter of a particle. Such conditions, close to plane-wave illumination, minimize any gradient contribution to the optical force field that turns out to be only determined by radiation pressure. 

Third, our setup can be operated in two illumination modes. In the single-beam mode, the laser beam illuminates the dispersion from one direction and pushes the particles along the optical axis $y$ --as described in Fig. \ref{fig1}(c). This mode is used for measuring radiation pressure forces, whose effects, along $y$, are decoupled from sedimentation and convection along the $z$ axis. The possibility to access a free diffusion regime is particularly important for carefully assessing the properties of the noise driving the colloidal system and for verifying its ergodic nature. But this regime is unreachable in the single-beam mode where the laser, even at its lowest intensity at the threshold of detection level, always pushes the particles. For this reason, we implement a dual-beam mode that consists in illuminating the dispersion from both sides along the $y$ axis. Because the laser beam is splitted by a polarizing beam splitter, the two counter-propagating beams are crossed polarized. They do not interfere and yield therefore a uniform intensity profile inside the cuvette. The intensities in each beam are carefully balanced so that radiation pressures coming from both sides of the cuvette can be perfectly compensated. In such conditions, the colloidal particles freely diffuse along the $y$ axis.

\section{Brownian dynamics under radiation pressure force field}

With gravity and convection acting along the $z$ axis only, the Brownian motion projected on the $y$ axis for each micron-sized colloidal particle can be described independently from these effects. Therefore, the motion of each particle, subjected inside the cell to the external optical force field $F$ directed along the $y$ axis and depending on both $y,z$ coordinates of the particle \footnote{The dependence on $x$ of the force field can be safely neglected considering that our imaging depth of field $\sim 8~\mu$m is much smaller than the beam waist $w_0\sim 65~\mu$m inside the cuvette. The particles imaged on our camera therefore diffuse along the $x$ axis within a constant laser intensity throughout the experiments}, is described by the simple overdamped Langevin equation:
\begin{equation}
\gamma \dot y_i(t) = F(y_i,z_i,t) + \sqrt{2k_{\rm B} T \gamma} \xi_i (t)   
\end{equation}
where $y_i(t)$ corresponds to the position of the $i^{\rm th}$ particle measured at a time $t$ along the $y$ axis, $k_{\rm B}$ the Boltzmann constant, $T$ the temperature of water, and $\gamma$ the Stokes friction drag. The stochastic Langevin force is modeled as a Wiener process that satisfies $\langle \xi_i(t)\rangle= 0$ and $\langle \xi_i(t) \xi_j(t') \rangle = \delta_{ij}\delta(t-t')$, with $\delta_{ij}=1$ if $i=j$ and $0$ if $i\neq j$, and where $\langle\cdots\rangle$ stands for an ensemble average performed over all the realizations of the stochastic variable $\xi_i(t)$. 

Experimentally, we acquire images at a given frame rate $f$ and thus implement a discrete version of the equation involving successive displacements $\Delta y_i(t_k) = y_i(t_{k+1}) - y_i(t_{k})$. The discrete Langevin equation reads as:
\begin{equation}
\gamma \frac{\Delta y_i(t_k)}{\Delta t} = F(y_i,z_i, t_k) + \sqrt{2k_{\rm B} T\gamma} \cdot \frac{w_i(t_k)}{\sqrt{\Delta t}} \label{langevin}
\end{equation}
where $t_k = k\cdot\Delta t$ with $\Delta t = 1/f$ and $k$ an integer number. Thermal fluctuations are set by $w_i(t_k)$ as a discrete random number, $\mathcal{N}(0,1)$ normally distributed, which satisfies $\langle w_i(t_k)\rangle = 0$ and $\langle w_i(t_k) w_j(t_l)\rangle = \delta _{ij} \delta_{kl}$. 

With a statistical distribution of displacements recorded over the ensemble $\{i\}$ of colloidal particles and for all times $\{k\}$ at a given laser intensity, we perform an ensemble averaging of Eq. (\ref{langevin})
\begin{equation}
\langle F(y_i,z_i,t_k)\rangle  = \gamma \frac{\langle \Delta y_i(t_k)\rangle }{\Delta t}  \label{shift}
\end{equation}
that allows to measure the strength of the force field, as soon as the Stokes friction drag is determined, as discussed in details below. 

We emphasize that this simple ensemble average relation relies on averaging out of the thermal fluctuation term in Eq. (\ref{langevin}). That $\langle w_i(t_k)\rangle=0$ in the ensemble of all displacements measured at different times and over different trajectories can only be true if one first confirms that all displacements are driven by the same source of thermal (white), stationary noise regardless of the chosen trajectory $i$ and the selected time $k$. 

\section{Thermal noise and stationarity}

\begin{figure}[htb]
  \centering{
    \includegraphics[width=0.95\linewidth]{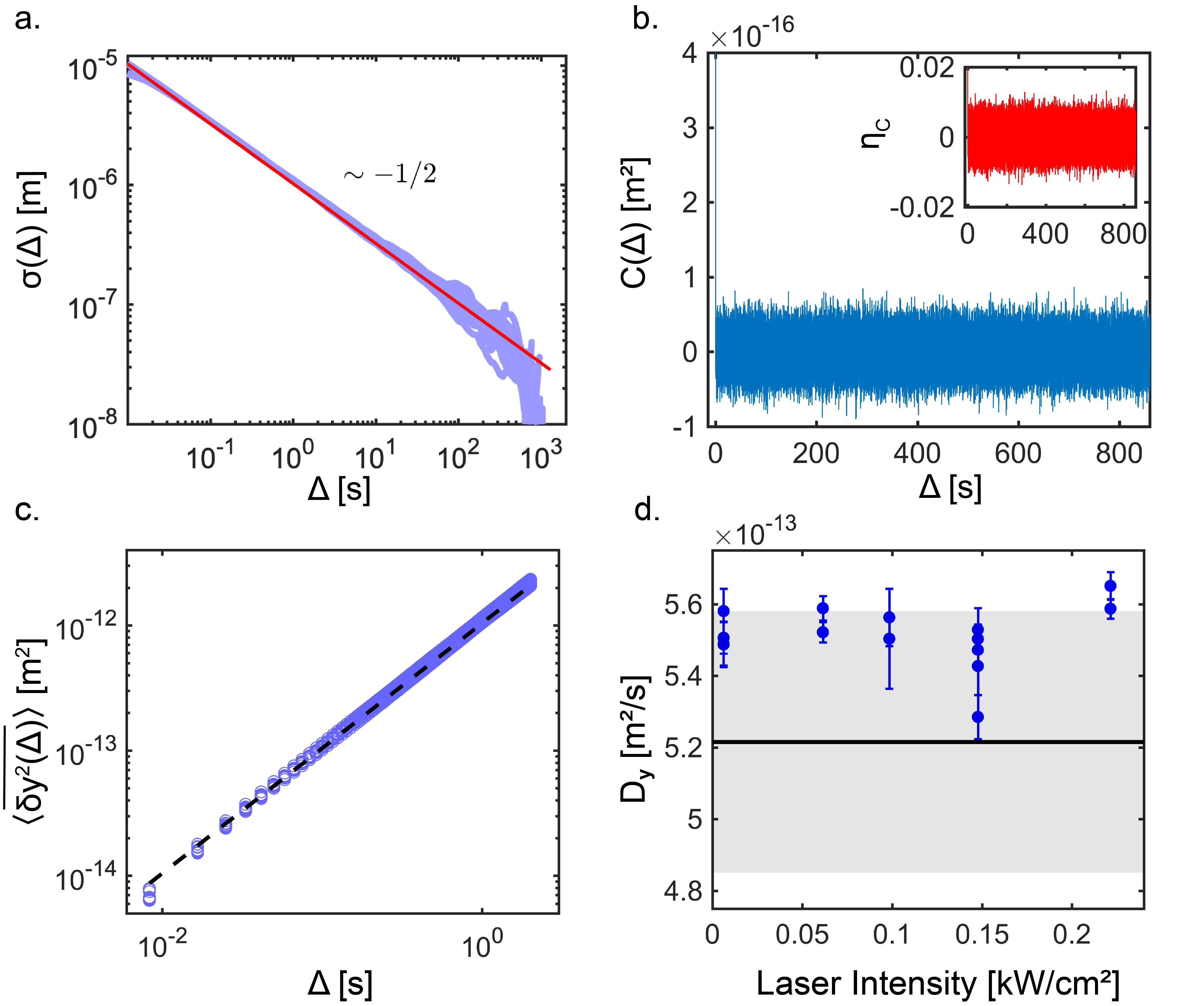}}
  \caption{(a) Allan variances calculated for 15 experiments performed in the dual-beam mode, using different laser intensities (from $0.006$ kW/cm$^2$ to $0.221$ kW/cm$^2$) as function of time lag $\Delta $. The close agreement with the Allan variance theoretically expected for a thermal white noise is shown with the red continuous line that corresponds to $\sigma(\Delta)=\sqrt{2k_{\rm B}T/(\gamma\Delta)}$ --see Appendix B for a demonstration. Note that the increased fluctuations in the Allan variances at large time lags only comes from a reduction in statistics. (b) The covariance and correlation of covariance evaluated from all the available displacements are plotted as a function of the time lag $\Delta$ to support the conclusion of the Allan variance. They both show the uncorrelated nature of the noise over a large temporal window of time lags.  (c) Experimental time ensemble average MSD measured along the $y$ axis as a function of time lag $\Delta$ for the different intensities given in panel (d). The superimposed dashed line shows the $2D\Delta$ MSD theoretically expected for a thermal white noise. (d) Diffusion coefficient $D_y$ (blue circles) extracted, for different laser intensities, from the different $y$-axis MSD shown in (c), by fitting the first 15 points of the MSD --following \cite{MichaletPRE2012} . Associated error bars at a 99.7\% confidence level on the MSD linear regression are displayed, taking into account tracking errors discussed in Appendix C. As seen on this panel, experimental values for $D_y$ are in a good agreement with the value $D=k_{\rm B} T /\gamma$ calculated at given temperature ($T = 298.65$K), viscosity ($\eta = 0.88\times 10^{-3}$ N.s.m$^{-2}$) and particle diameter ($d = 940nm$). The gray zone corresponds to uncertainty in the diffusion coefficient coming from errors in temperature ($\delta T = \pm 1.5$K), viscosity ($\delta \eta = \pm0.01$ N.s.m$^{-2}$) and particle diameter ($\delta d = \pm50$nm).}    
\label{fig2}
\end{figure}

In order to verify the thermal and stationary properties of the noise at play in our colloidal system, we illuminate the colloidal dispersion in the dual-beam mode corresponding to two external force fields that compensate each other. Under such conditions, we perform an Allan variance-based analysis on a reconstructed single long trajectory composed of all concatenated displacements $\{\Delta y_i (t_k)\}$ over the ensemble of trajectories $\{i\}$. This calculation, detailed in Appendix B, is performed over 15 different experiments and the result is displayed in Fig. \ref{fig2}(a). The Allan variance clearly shows, in log-log scale, the $-1/2$ exponent expected for a system driven by a white thermal noise  \cite{Allan1966}. Remarkably, this behavior spans ca. 3 decades of time lag $\Delta$, showing no drift in the experimental thermal noise and thereby indicating that one can exploit the full ensemble of displacements indeed.

As reminded in Appendix B, the Allan variance is related to the displacement covariance $C(\Delta, t) = \langle \Delta y_{\Delta  + t}  \Delta y_t  \rangle$ which can also be directly evaluated from all the displacement available in the concatenated trajectory. The result is displayed in Fig. \ref{fig2}(b) showing that $C(\Delta, t)$  fluctuates around a zero mean for all time lags $\Delta >0$, as expected from the white noise Allan deviation. The degree of stochastic independence of each successive steps can be quantified using the correlation of covariance $\eta = {C(\Delta)}/{\sigma_{\Delta y}^2}$ which, as expected again for a white noise, fluctuates around a zero mean with an amplitude much smaller than $1$. 

It also interesting to look at the time ensemble average mean square displacement (MSD) $\langle \overline{\delta y_i ^2(\Delta)}\rangle$ evaluated by, first, averaging over their duration $\mathcal{T}_i$ every single trajectory $i$ MSD
\begin{equation}
\overline{\delta y_i ^2 (\Delta)} = \frac{1}{\mathcal{T}_i-\Delta}\int_0^{\mathcal{T}_i-\Delta} (y_i(t'+\Delta) - y_i(t'))^2 dt'  ,
\label{TAMSD}
\end{equation}
and, then, taking the mean over the ensemble $\{i\}$ of trajectories of such time averages. 

The results, evaluated under the same dual-beam mode of illumination for different laser intensities in each beam, are displayed in Fig. \ref{fig2}(c). The clear linearity of $\langle \overline{\delta y_i ^2(\Delta)}\rangle$, for all time lags $\Delta$ corresponds to two important characteristics of the stochastic process driving the particles: first, a zero mean (linearity) and second, a slope $2D$ that measures the diffusion coefficient $D$ of the free Brownian motion along the $y$ axis. As shown in Fig. \ref{fig2}(d) for different laser intensities injected in each path of the dual-beam mode, this measurement falls in good agreement with the value $D=k_{\rm B}T/\gamma$ expected from the Langevin equation, including systematic tracking errors, temperature value and particle size dispersion errors, as discussed in details in Appendices C and D. 

The zero mean, fixed variance and covariance independent of the time lag, all together manifest the white stationary character of the noise driving our experiments, in full agreement with the stochastic description of the Langevin force in Eq. (\ref{langevin}).

\section{Ergodicity}

This analysis implies that all the successive displacements drawn from different single trajectories at different times within the concatenated trajectory are driven by the same white noise. Under such conditions, the colloidal dispersion must behave like an ergodic system, at the level of which it is possible to collect displacement values acquired from different $i$ trajectories at different $k$ times and to perform large time ensemble averages necessary to evaluate Eq. (\ref{shift}) with great precision. In the context of high resolution force measurements therefore, ergodicity is an important property to verify.

\begin{figure}[htb]
  \centering{
    \includegraphics[width=0.8\linewidth]{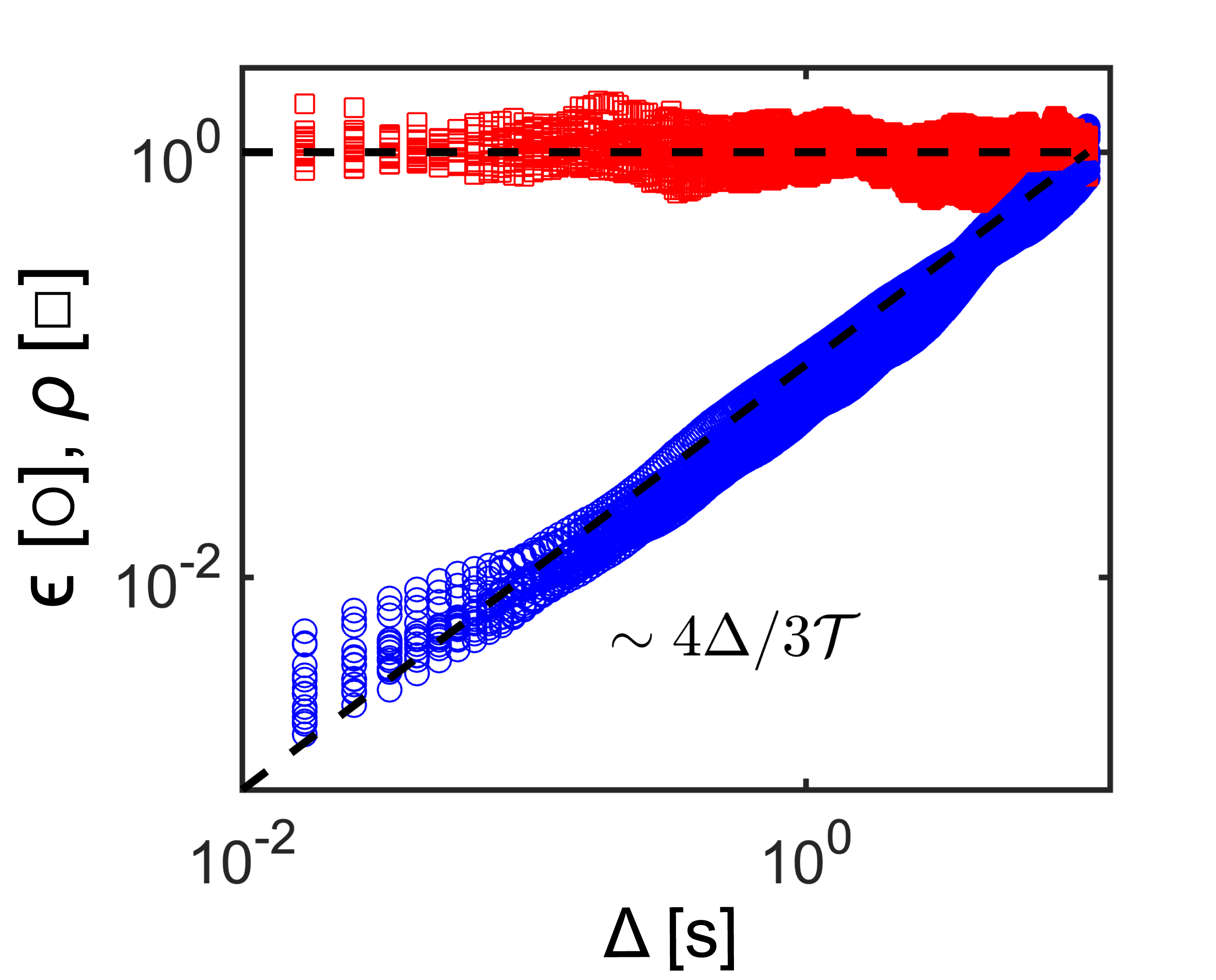}}
  \caption{Ergodic parameters $\epsilon$ (blue circles) displayed, for all time lags, for all the 15 experiments gathered in Fig. \ref{fig2}(c) and (d). These results are all corrected for tracking errors that particularly impact the data at small time lags, as discussed in details in Appendix F. As clearly seen, $\epsilon(\Delta)$ follows the evolution expected for an ergodic free Brownian motion. All ratio $\rho(\Delta) = \left< \overline{\delta y^2 (\Delta)} \right> / \left< y^2 (\Delta) \right>$) (red squares) evaluated for the same experiments remain, as expected, close to $1$ for all time lags $\Delta$. We note that both quantities establish the necessary and sufficient conditions for ergodicity discussed in \cite{Barkai2014}.
}
  \label{fig3}
\end{figure}

Stationarity implies that the ensemble average mean-square displacement (MSD) 
\begin{equation}
\langle \delta y^2 (\Delta) \rangle = \frac{1}{N}\sum_{i} (y_i(t+\Delta) - y_i(t))^2
\end{equation}
is independent from the choice of the initial time $t$. As a consequence, $\langle \delta y^2 (\Delta) \rangle$ equals its time average
that simply corresponds to the ensemble mean $ \langle \overline{\delta y_i ^2(\Delta)}  \rangle$ of single $\{i\}$ trajectory time average MSD given by Eq. (\ref{TAMSD}). This property is clearly seen in Fig. \ref{fig3} with a ratio $\rho(\Delta) = \langle\overline{\delta y_i ^2(\Delta)} \rangle / \langle \delta y^2 (\Delta) \rangle$ equal to one for all time lags. 

Ergodicity \textit{per se} then requires any single-trajectory time average MSD $\overline{\delta y_i ^2(\Delta)}$ drawn from the trajectory ensemble $\{i\}$ to become equal to the ensemble mean of such time average MSD 
\begin{equation}
\lim_{\mathcal{T}/\Delta \rightarrow\infty} \overline{\delta y_i ^2(\Delta)} = \langle \overline{\delta y_i ^2 (\Delta)} \rangle \label{ergo1}
\end{equation}
for a sufficiently long averaging time $\mathcal{T}$. This time corresponds to the full duration of the shortest trajectory recorded and hence defines the duration on which all trajectories are limited or possibly, if long enough, subdivided. 

It is clear that this sufficient condition for ergodicity can be quantified by looking at the evolution of the statistical distribution of the single-trajectory MSD $\overline{\delta y_i ^2(\Delta)}$ which, according to Eq. (\ref{ergo1}), is expected to reach a mean value equal to $\langle \overline{\delta y_i ^2(\Delta)} \rangle$ with a variance $\sigma^2 \left( \overline{\delta y_i^2 (\Delta)}  \right)$ decreasing as $\mathcal{T}/\Delta$ increases. 
Accordingly, and following \cite{Barkai2014}, we introduce the ergodic parameter
\begin{equation}
\epsilon (\Delta)= \frac{\sigma^2 \left( \overline{\delta y_i^2 (\Delta)}  \right)}{\langle \overline{\delta y_i ^2(\Delta)} \rangle^2} = \frac{\langle \left( \overline{\delta y_i^2 (\Delta)} \right) ^2 \rangle }{\langle \overline{\delta y_i ^2(\Delta)} \rangle^2} - 1  \label{eqEB}
\end{equation}
that can be explicitly evaluated for the case of the one-dimensional free Brownian motion expected to take place in our experiments along the optical axis $y$ in the dual-beam mode. The calculation is detailed in Appendix E and leads to the simple ergodic law
\begin{equation}
\lim_{\mathcal{T}/\Delta \rightarrow \infty} \epsilon(\Delta) = \frac{4\Delta}{3\mathcal{T}}.  \label{ergo}
\end{equation}

As shown in Fig. \ref{fig3}, this law is clearly followed by our data. We stress in Appendix F that apparent deviations of $\epsilon$ from the ergodic law at short time lags are in fact induced by tracking errors on displacement measurements. Once these errors are accounted for, our measurements eventually agree well with Eq. (\ref{ergo}), clearly demonstrating the ergodic nature of our colloidal dispersion.

\section{Radiation pressure force measurement and profile reconstruction}

White noise stationarity and ergodicity ensure that Eq. (\ref{shift}) is experimentally valid for our system. It can therefore be evaluated by collecting all displacements in order to perform the ensemble averages. Because radiation pressure is determined by fixed laser intensities, and therefore constant in time, $\langle F(y_i,z_i,t_k)\rangle=\langle F(y_i,z_i)\rangle$ in Eq. (\ref{shift}). In addition, the Rayleigh range of our Gaussian illumination mode is large compared with the imaging region-of-interest. This gives a radiation pressure force field invariant along the $y$ axis and only dependent on $z$ with an expected Gaussian profile centered on a maximal force $F_0$
\begin{equation}
F_{y}(z) = F_{0} \exp\left(-\frac{2 (z-z_0)^2}{w_0^2} \right),  \label{profile}
\end{equation}
where the central position $z_0$ of the laser beam and its waist $w_0$ are determined optically with a high level of precision (see below). Under such conditions, $\langle F_i(y_i,z_i)\rangle$ can be evaluated as an average performed on the successive $z$ positions that all diffusing colloids cross within the laser beam while being pushed along the $y$ axis. Experimentally therefore, we record the $(y,z)$ positions of each colloid simultaneously and build two statistical ensembles gathering, one, the successive $n^{\rm th}$ $\Delta y_n=y_{n+1}-y_n$ displacements measured along the $y$ axis, and the other, the corresponding $z_n$ positions of the colloids on the $z$ axis recorded for the $\Delta y_n$ displacement. We will note $N$ the total number of such displacements available from all the trajectories and at all times in one experiment. 

\begin{figure}[htb]
  \centering{
    \includegraphics[width=0.8\linewidth]{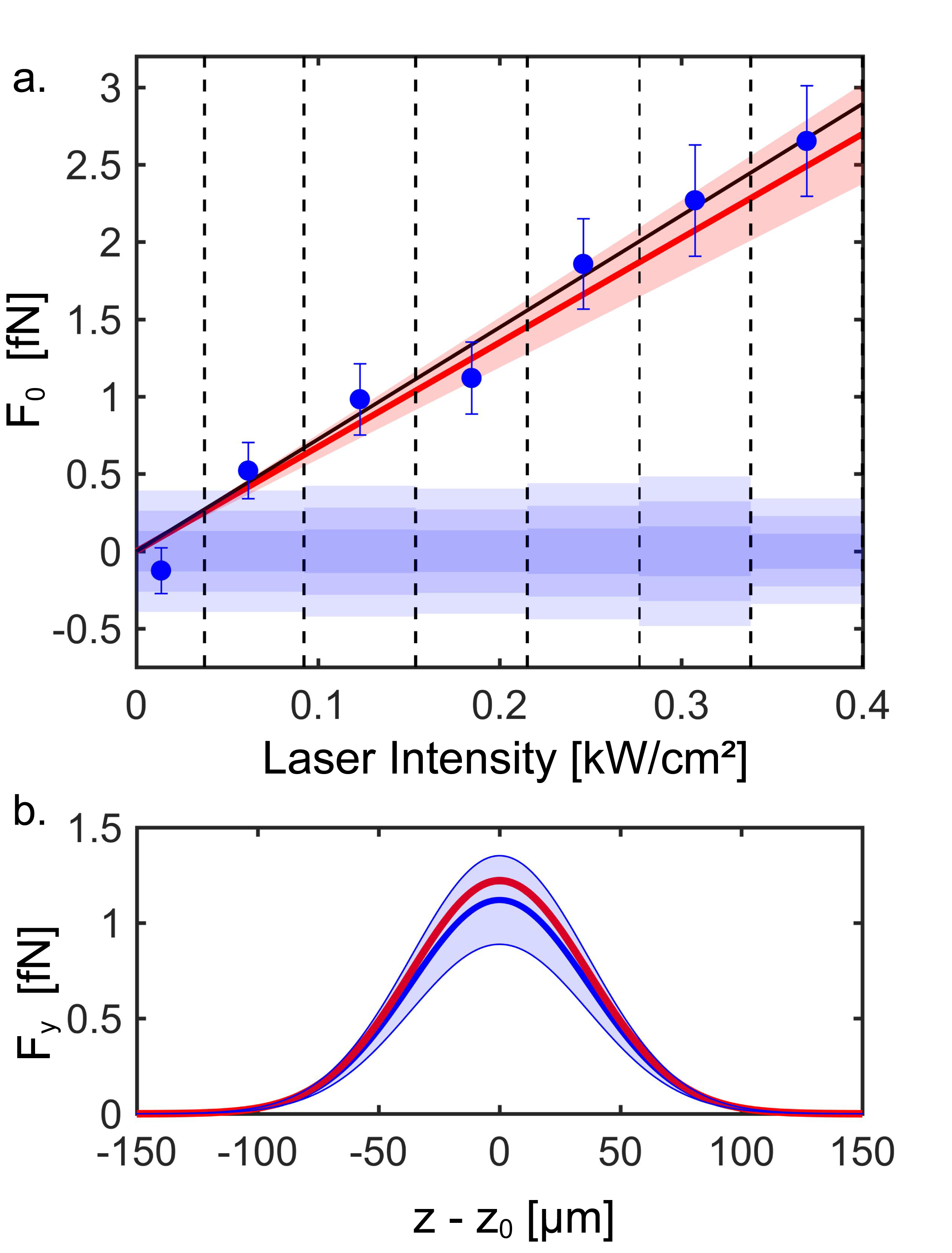}}
  \caption{(a) Maximal radiation pressure force estimator $F_0$ evaluated as a function of different laser intensities in the single-beam configuration of the experiment sketched in Fig. \ref{fig1}. The estimators (blue points) show the expected linear dependence on the laser power and the continuous black line is a linear fitting performed as a reference. The error bars on the estimators are discussed in the Appendix G. The Mie calculation (red line) done for the conditions of the experiment (plane wave approximation, spherical dielectric -melamine- particles) is also shown, with its uncertainty (light red surface) detailed in the Appendix H. The blue surfaces centered around the $F_0=0$ value display the thermal limit at 1,2, and 3 successive confidence levels. Each experiment corresponds to a given $N$ value, hence a given thermal limit in the force detection. (b) With the estimator $F_0$, the entire radiation pressure force profile can be reconstructed. A $F_y(z)$ profile is shown as an example, corresponding to a $0.18$ kW$\cdot {\rm cm}^{-2}$ laser intensity (i.e. $15$ mW total laser power). The blue surface describes the uncertainty associated with the force profile reconstruction --see main text.}
  \label{fig4}
\end{figure}

Such two statistical ensembles give the possibility to estimate the maximal force $F_0$ from the agreement between the ensemble averages performed respectively over the ${\{z_n\}}$ and ${\{\Delta y_n\}}$ ensembles
\begin{equation}
F_{0} \langle\exp\left(-\frac{2 (z_n-z_0)^2}{w_0^2} \right) \rangle \equiv \frac{\gamma}{\Delta t} \langle \Delta y_n \rangle.   \label{estim}
\end{equation}
The estimators of $F_0$ evaluated for different laser intensities in the single-beam configuration are presented in Fig. \ref{fig4}(a), including $\pm\delta F_0$ error bars associated with the different sources of errors described in detail in the Appendix G. The results, linearly dependent on the laser intensity, are in good agreement with radiation pressure force values expected from a Mie calculation \cite{CanaguierPRA2014}. 

Once the estimator $F_0$ is determined, the full force profile (\ref{profile}) can be reconstructed, as shown in Fig. \ref{fig4}(b) for a given laser intensity. Associated uncertainties here include the residual errors in determining optically the $(z_0,w_0)$ values. The waist $w_0$ is measured by imaging, with a camera facing the laser, the transverse intensity of the laser beam. Analyzing the profile with a Gaussian model, the waist value ($65~\mu$m) is estimated with a relative error of $4.9\%$. The beam center $z_0$ is determined by superposing all the frames (gathering $\sim 3\times 10^6$ fluorescent spots) from all the experiments, by analyzing the upper and lower limits and the distribution of all points in the imaging field of view, leading to estimate the center position $z_0$ with a relative error of $0.2\%$.

A force resolution level of our thermally limited system can be derived, for each experiment providing $N$ displacements, as the minimal measurable force 
\begin{equation}
\langle F \rangle_{min} = m \cdot \frac{\gamma}{\Delta t} \sigma \left( \langle \Delta y_n \rangle \right)    
\end{equation}
where $m$ fixes the confidence interval chosen, and  $\sigma \left( \langle \Delta y_n \rangle \right) = \sigma \left( \Delta y_n \right) /\sqrt{N}$ is the standard error evaluated from the variance $\sigma \left( \Delta y_n \right)$ of the displacement ensemble $\{\Delta y_n\}$. This variance can be evaluated directly from Eq. (\ref{langevin}) in the absence of any force field as
\begin{equation}
\sigma (\Delta y_n) = \sqrt{\frac{2k_BT\Delta t}{\gamma}},
\end{equation}
giving a resolution level of our force measurement as:
\begin{equation}
\langle F\rangle_{min} = \sqrt{2k_B T \gamma} \frac{m}{\sqrt{N\Delta t}}.
\end{equation}

As clear from this relation, the resolution can be improved by increasing the size of the statistical ensemble (i.e. the number $N$ of displacements) within the thermal, stationary and ergodic limits set in Sec. B. For our experiments, the thermally limited force sensitivity is $\sqrt{2k_B T \gamma}\simeq 8.56$ N$/\sqrt{\rm Hz}$. Within such limits, a typical experiment yields $N \sim 2 \times 10^5$ successive displacements, which corresponds to a remarkable $\sim 0.3$ fN resolution level (within a $m=3$ $99.7\%$ confidence interval). Such levels of resolution are displayed in Fig. \ref{fig4}(a) and they vary just as $N$ depends on the experiment performed at a given laser intensity. These data clearly show that in the low laser power regime, radiation pressure can be determined down to the thermal limit, with a $\sim0.5$ fN force actually measured at a $99.7\%$ confidence level.

\begin{figure}[h]
  \centering{
    \includegraphics[width=0.7\linewidth]{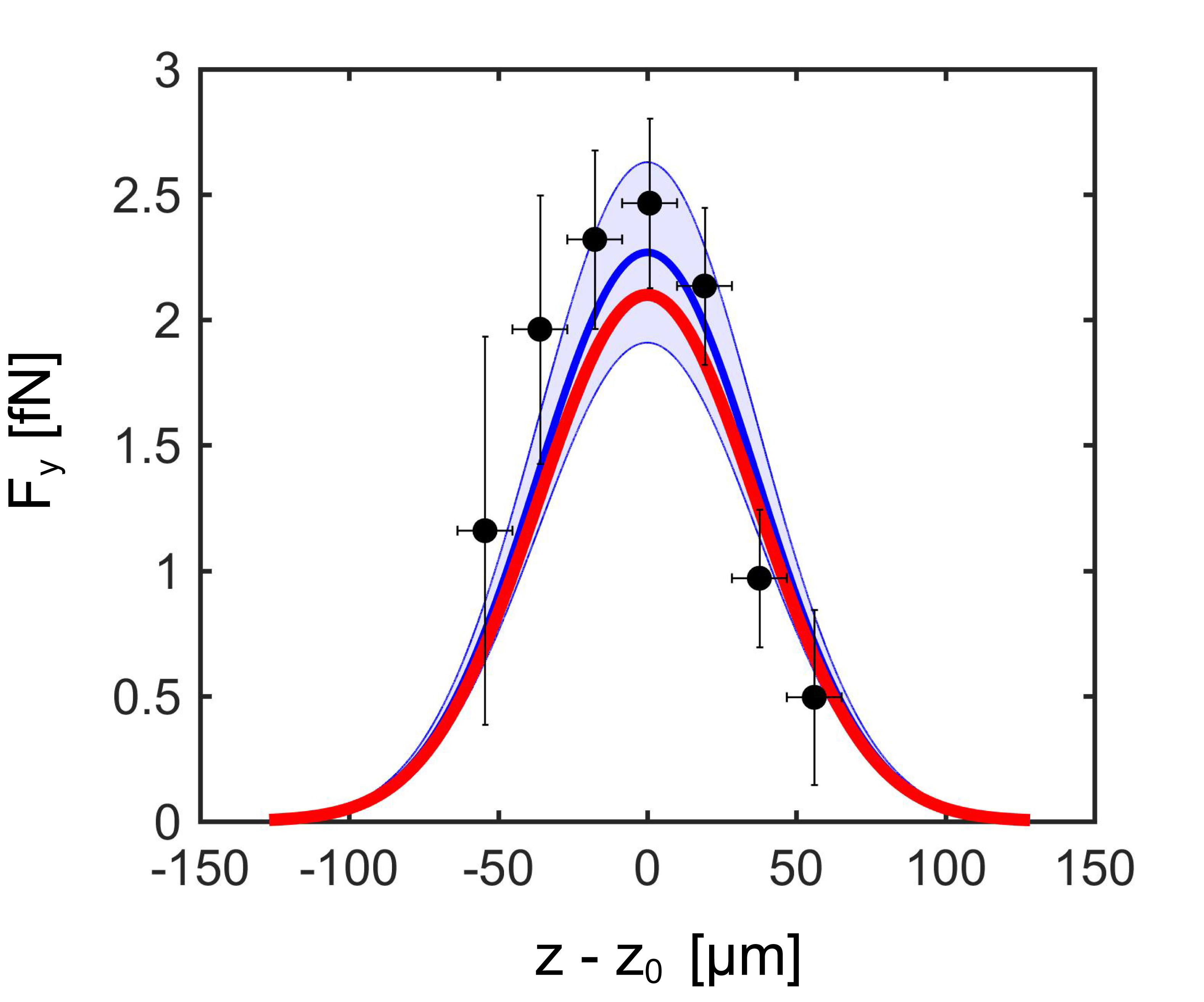}}
  \caption{Radiation pressure forces measured zone-by-zone (in successive layers of identical thicknesses $18.4~\mu$m corresponding to the horizontal widths of all error bars) along the $y$ optical axis for a laser intensity of $0.31$ kW/cm$^{2}$ --corresponding to a total laser power of 25 mW. These measurements, together with their associated error bars --following the analysis presented in Appendix G-- are compared to the reconstructed profile using our statistical ensemble approach (blue line) including uncertainties (blue surface) and Mie calculations (red line). The number statistics for each layer of data (from left to right: 8633, 17785, 41573, 45241, 51718, 67707 and 43249 recorded displacements) impacts the resolution, as clearly seen from the variable vertical widths of the corresponding error bars. }
  \label{figS4}
\end{figure}

Finally, we exploit the fact that video-tracking microscopy also gives the possibility to measure the external optical force field directly within limited horizontal layers chosen from both sides of the optical axis of the illumination beam. We define $18.4~\mu$m thick layers at given $z$ positions and within which we collect all displacements along the $y$ direction. We use again Eq. (\ref{shift}) inside such selected layers to give a measurement of the average radiation pressure force zone-by-zone. The results are reported in Fig. \ref{figS4} and falls in very good agreement with the reconstruction method.

\section{Conclusions}

Our results demonstrate that colloidal dispersions can be used as highly sensitive Brownian probes for measuring external force fields. Because such systems remain thermally limited, with stationary and ergodic dynamics, large statistical ensembles mixing particle displacements recorded from different trajectories and times become available and thus make sub-fN force resolution levels accessible. The main experimental uncertainty stems in our work from the particle size dispersion in the colloidal dispersion commercially available. But this source of systematic error could be reduced working with better monodisperse colloidal dispersions. Our method of statistical reconstruction of the force profile, here demonstrated in two-dimensions for a non-conservative optical force field, can obviously be extended in three-dimensions, involving efficient and available techniques in three-dimensional tracking \cite{MartinezMarradesOptX2014,VerrierApplOpt2015}. Validating colloidal systems for ultrasensitive force detection strategies, our work opens interesting perspectives. For instance, colloids can be involved in reconstructing complex topological force fields, in particular in the near field where momentum exchanges are enhanced by near-field inhomogeneities \cite{Cuche2012,CanaguierPRA2013}. As recently proposed, colloidal systems can also become pertinent tools for weak force measurements in the context of Casimir physics \cite{MaiaNeto2015}. The advent of nano- and micro-structured and functionalized colloids \cite{Norris2014,Marzan2016,Kotov2017} can lead to new types of dynamical responses to external fields, as exemplified with chiral optical forces \cite{CanaguierNJP2013,Cameron2014,Chan2014,Dionne2017,BrasseletPRL2019,BrasseletPRAppl2019}. The outstanding stability of the statistical properties of our system offers new possibilities for deciphering non-trivial force fields at a genuine sub-fN resolution level. We finally stress that our approach is also relevant for weak force experiments that do not necessarily involve optical fields, experiments found for instance in the burgeoning field of mechanical chiral resolution \cite{Marichez2019}.

\section{Acknowledgments}

This work was supported in part by Agence Nationale de la Recherche (ANR), France, ANR Equipex Union (Grant No. ANR-10-EQPX-52-01), the Labex NIE projects (Grant No. ANR-11-LABX-0058-NIE), and USIAS within the Investissements d'Avenir program (Grant No. ANR-10-IDEX-0002-02). 

\section{Appendix A: Sample preparation}

The samples are prepared from an initial dispersion ($2.5\%$ mass-volume ratio) of melamine micro-particles of diameter $d = 0.940\pm0.05~\mu$m purchased from Micro-Particles GmbH, weakly doped with a fluorescent dye for a most efficient detection in water. We dilute the dispersion $\sim10^4 \times$ with ultra-pure water and fill a cuvette with the colloidal dispersion to a $1.3\times 10^{-6}$ low-volume fraction. The dimensions of the cuvette --see Fig. \ref{fig1} in the main text-- are large enough so that within imaging region-of-interests, boundary wall effects can be safely neglected. The filled cuvette is covered and sealed with vacuum grease in order to prevent water evaporation and to isolate the fluid from other environmental influence. The cuvette and its cover are exposed at least one hour to UV light in order to ensure the absence of any bacterial contaminant. The sample is grounded in order to remove any electrostatic charge on the surface of the cuvette. Before performing our experiments, we leave for about $1$ hour the sample relaxing in its holder until well thermalized with the environment. 

One sees in our system some convective drag effects at play, induced by the illuminating laser depending on its intensity. These effects can be simply understood. When the laser is turned on, the water is heated. Although this effect is minute, it is sufficient for reducing the density of water within the laser beam. As a consequence, buoyancy driven flows are induced that can eventually drag the particles upwards at the highest laser intensities. This is what is seen, for instance, in Fig. \ref{fig1} (c) where the convection induced at high laser intensity drags the melamine spheres against sedimentation. The large dimensions of the cuvette insure that such a convective drag is strictly laminar and performed along the $z>0$ axis. In such conditions, both sedimentation and convective effects are perfectly decoupled from the diffusive dynamics performed on the optical $y$ axis, along which external force fields are measured.

It is interesting to note that for smaller laser intensities, the drag resulting from sedimentation and convection is directed downwards. This means that it is actually possible to find a condition of illumination where sedimentation and convective drags can compensate each other. In the dual-beam illumination mode, this condition corresponds to a laser intensity of $0.148$ kW/cm$^{2}$ in each beam. Remarkably then, the Brownian motion of the ensemble becomes totally free in three dimensions within the imaging field of view.

\section{Appendix B: Allan variance analysis of the noise}

The Allan variance is a statistical tool particularly relevant for quantifying noise sources because of its direct relation with noise power spectral densities \cite{Allan1966,Barnes1971} . It has recently been used in the context of optical trapping in order to assess limits of stability of traps and to provide optimal measurement bandwidths \cite{Gibson2008,Oddershede2009,SchnoeringPRAppl2019}. 

We first explain how the Allan variance is measured in our experiments, performed in the dual-beam mode with two laser beams of equal intensity illuminating the dispersion. This leads to a precise compensation of external radiation pressure forces --i.e. to free Brownian motion along the $y$ axis, described by the simple Langevin equation
\begin{equation}
\Delta y_i(t_k) = \sqrt{\frac{2k_B T \Delta t}{\gamma }} \cdot w_i(t_k),
\end{equation}
only driven by the noise $w_i(t_k)$. In such conditions, the successive displacements $\Delta y_i(t_k)$ recorded from all the trajectories (over the shortest observation time lag $\Delta t$) are concatenated one-by-one in order to form a single time series $\Omega(\mathcal{\tau})$ of displacements with a total time $\mathcal{T}=N\Delta t > \tau$. 

The Allan variance
\begin{equation}
\sigma^2(\tau) = \frac{1}{2}\langle\Delta \overline{\Omega}^2(\tau) \rangle
\end{equation}
is calculated on this long time series using successive and non-overlapping series of shorter duration $\tau$ with $\Delta \overline{\Omega}(\tau) = \overline{\Omega}_{k+n}(\tau) - \overline{\Omega}_k(\tau)$ and for which time average values $\overline{\Omega}_k(\tau)$ of $\Omega(\tau)$ are defined as:
\begin{equation}
\overline{\Omega}_k(\tau) = \frac{1}{\tau} \int_{t_k}^{t_k+\tau} \Omega(t^\prime) dt^\prime \label{tav}
\end{equation}
with $t_k = k\Delta t$ and $\tau = n\Delta t$. Here, $\langle \cdots \rangle$ defines an average performed over all accessible series, implying that $n< N/2$ and $k\leq N-2n$. The Allan variance is then explicitly calculated as:
\begin{equation}
\sigma^2(\tau) = \frac{1}{2(N-2n)}\sum^{N-2n}_{k=1}\left( \overline{\Omega}_{k+n}(\tau) - \overline{\Omega}_{k}(\tau) \right)^2.  \label{AllanVar}
\end{equation}

The connection to noise power spectral density can be seen by defining an intermediate estimator $\theta$
\begin{equation}
\theta(\tau) = \int_0^{\tau}\Omega(t^\prime)dt^\prime
\end{equation}
that corresponds in \cite{Allan1966} to instantaneous phase fluctuations for $\Omega(\tau)$ corresponding to instantaneous frequency fluctuations. With this estimator, the Allan variance can be written as
\begin{equation}
\sigma^2(\tau) = \frac{1}{2\tau^2}\langle\left( \theta((i+2)\tau) - 2\theta((i+1)\tau) + \theta(i\tau)\right)^2\rangle, \label{Allanthe}
\end{equation}
which gives a sum of different covariance function $C_{\theta}(\Delta) = \langle \theta(\tau+\Delta) \theta(\tau) \rangle$ related to a power spectrum densitiy (PSD) as
\begin{equation}
C_{\theta}(\Delta) = \frac{1}{2\pi} \int_{-\infty}^{+\infty} S_{\theta}(\omega) e^{i\omega \Delta}d\omega.
\end{equation}
Eq. (\ref{Allanthe}) can thus be simply expressed as
\begin{equation}
\begin{aligned}
\sigma^2(\tau) &=\frac{1}{4\pi\tau^2}\int_{-\infty}^{+\infty} d\omega S_{\theta}(\omega)\big[ 6-2(e^{i\omega\tau}+ e^{-i\omega\tau})\\
&~~~ + (e^{2i\omega\tau}+ e^{-2i\omega\tau}) - 2(e^{i\omega\tau}+ e^{-i\omega\tau}) \big]\\
&= \frac{1}{4\pi\tau^2}\int_{-\infty}^{+\infty} d\omega S_{\theta}(\omega) \big[ 6 - 8\cos(\omega\tau) + 2\cos(2\omega\tau)\big]\\
&= \frac{4}{\pi\tau^2}\int_{-\infty}^{+\infty} d\omega S_{\theta}(\omega) \sin^4\left(\frac{\omega\tau}{2}\right).
\end{aligned}
\end{equation}
Considering the relation $\omega^2 S_{\theta}(\omega)=S_{\Omega}(\omega)$ between the PSD of the noise and the PSD of its estimator, we have
\begin{equation}
\sigma^2(\tau) = \frac{4}{\pi \tau^2}\int_{-\infty}^{+\infty} \frac{d\omega}{\omega^2} S_{\Omega}(\omega) \sin^4\left(\frac{\omega \tau}{2} \right),  \label{AllanPSD}
\end{equation}
as the important relation that shows why the Allan variance is an appropriate tool for quantifying all kind of noise sources characterized by their respective PSD. 

Let us evaluate the Allan variance for the case of a white noise $\Omega(t)=w(t)$ characterized by the well-known covariance $\langle w(t)w(0)\rangle = h_0 \delta (t)$ and the associated PSD $S_w (\omega)=h_0$. For such a PSD, we can directly evaluate Eq. (\ref{AllanPSD}) to find that $\sigma^2(\tau) = \frac{h_0}{\tau}$. For the \textit{thermal} white noise driving our experiments, $h_0 = 2k_{\rm B}T/\gamma$, leading to
\begin{equation}
\sigma(\tau) = \sqrt{\frac{2k_{\rm B}T}{\gamma \tau}}.
\end{equation}
This type of Allan variance is clearly verified in our experiments (Fig. \ref{fig2} (a), main text) with a  linear evolution observed in the log-log scale with a $-1/2$ slope and an amplitude fixed at $2k_{\rm B}T/\gamma$.

Experimentally, the Allan variance is calculated from a discretized version of the estimator
\begin{equation}
\theta_k = \Delta t\sum_{j = 1}^{k} \Omega_j
\end{equation}
where $\theta_k = \theta (k\Delta t)$ and $\Omega_j=\Omega(j\Delta t)$. The time averages of $\Omega$ --Eq.(\ref{tav})-- performed over the temporal length of the trajectory $\tau =n\Delta t$ for two successive and non-overlapping displacement samples, is then simply expressed as:
\begin{equation}
\overline\Omega _k(\tau) = \frac{\theta_{k+n} - \theta_{k}}{\tau}
\end{equation}
\begin{equation}
\overline\Omega _{k+n}(\tau) = \frac{\theta_{k+2n} - \theta_{k+n}}{\tau}.
\end{equation}
Following Eq. (\ref{AllanVar}), we therefore calculate in this work the Allan variance as:
\begin{equation}
\sigma^2(\tau) = \frac{1}{2(N-2n)\tau^2}\sum_{k = 1}^{N-2n} (\theta_{k+2n} - 2\theta_{k+n} + \theta_k)^2.
\label{Allancal}
\end{equation}

\section{Appendix C: Tracking error data correction}

As discussed in \cite{Michalet2012}, recording the trajectory of a diffusing particle under the microscope is accompanied by tracking errors that can affect the Brownian statistical analysis. Such errors combine localization errors for single-shot positional measurements and blurring effects caused by the finite exposure time $t_E$ of such a measurement over which the particle keeps diffusing with a diffusion coefficient $D$. For free Brownian motion, the one-dimensional time average mean-square displacement (MSD) acquired along a given axis (the $y$ axis, following our experimental setup) has to be corrected. In order to include these effects, the MSD writes as:
\begin{equation}
\langle \overline{\delta y^2(\Delta)} \rangle= 2D\cdot\left(\Delta - \frac{t_E}{3} \right) + 2\sigma^2
\end{equation}
where $t_E$ is the exposure time of the camera and $\sigma$ is the dynamic localization error. This error
\begin{equation}
\sigma = \sigma_0 \left( 1 + \frac{Dt_E}{s_0^2} \right)^{1/2} \label{error}
\end{equation}
is related to the static localization error $\sigma_0$ and the full-width at half maximum $s_0$ of the Gaussian profile used to approximate the microscope point-spread function. 

For our experimental setup, we have: $\sigma_0\sim 15$ nm and $s_0\sim 1.5~\mu$m, extending over $3$ pixels of the CCD camera. Our exposure time is fixed at $t_E=7$ ms. Using the expected value of $D$ in water at room temperature for a particle of diameter $940$ nm, these values implies that $s_0^2 \gg Dt_E$, so that $\sigma \sim \sigma_0$.

Therefore, the single-particle time average MSD $\overline{\delta y_i^2(\Delta)}_{exp}$ experimentally acquired 
\begin{equation}
\overline{\delta y_i^2(\Delta)}_{exp} = \overline{\delta y_i^2(\Delta)}_0  + \varepsilon_i  \label{deltaExp}
\end{equation}
departs from the real MSD $\overline{\delta y_i^2(\Delta)}_0$ by a random localization error $\varepsilon_i$. At the level of a colloidal dispersion, we perform an additional ensemble averaging giving 
\begin{equation}
\langle \overline{\delta y_i^2(\Delta)} \rangle_{exp} = \langle \overline{\delta y_i^2(\Delta)} \rangle_{0} + b  \label{track}
\end{equation}
with $\langle \overline{\delta y_i^2(\Delta)} \rangle_{0}=2D\Delta$ and $b=\langle \varepsilon_i \rangle=2\sigma_0^2 -2Dt_E/3$. Within our experimental conditions, this error parameter is estimated to be $b\sim -2.6\times 10^{-3}~\mu{\rm m}^2$, negative as expected looking at the experimental MSD data point at small $\Delta$.

Such corrections impact the determination of the experimental MSD under an external force field, hence the diffusion coefficient, and the determination of the ergodicity parameter $\epsilon$ that can gives the impression, if not corrected, to display signatures of ergodicity breaking at short time scales, as discussed in Sec. F in details. 

We can also get an estimate of the tracking error by linearly fitting the 15 experimental time ensemble average MSD. We obtain an average error parameter $b_{\rm fit}\sim -2.8\times 10^{-3}~\mu{\rm m}^2$ including a standard deviation of $\sim 0.4\times 10^{-3}~\mu{\rm m}^2$, showing that $b_{\rm fit}$ is in good agreement with the estimated value $b$.

\section{Appendix D: Determination of diffusion coefficients}

Determining diffusion coefficients is central to our experiments since it allows to extract a value for the Stokes friction drag $\gamma$ once the temperature of the system is known. Under an external force field, the diffusion coefficient is related to the experimental variance
\begin{equation}
\sigma^2(r(\Delta+t) - r(t)) = 2D\Delta +b
\label{sigfit}
\end{equation}
given by the statistical ensemble of displacements measured along one $r$ axis with a time difference $\Delta$, including the tracking error $b$ discussed above in Eq. (\ref{track}).

\begin{figure}[htb]
  \centering{
    \includegraphics[width=0.97\linewidth]{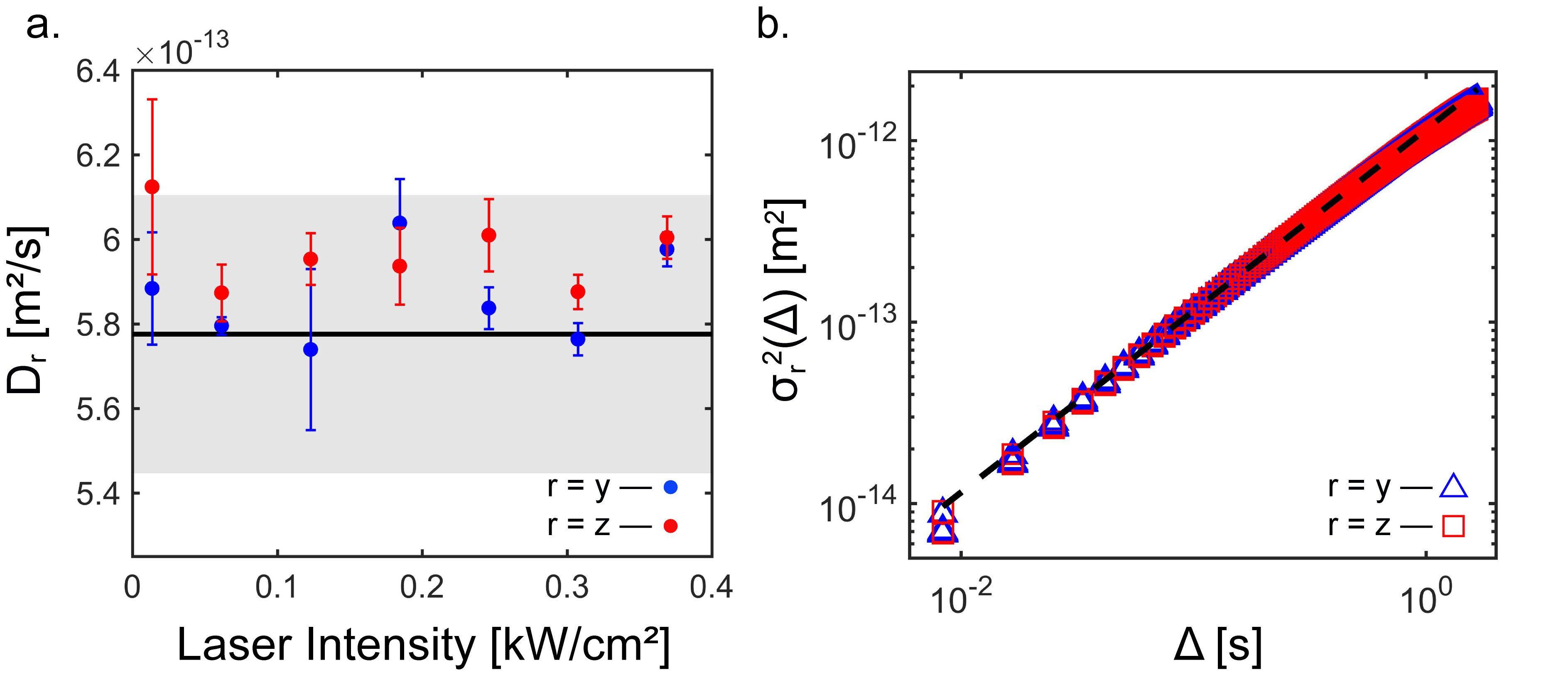}}
  \caption{(a) Diffusion coefficients $D_r$ measured along the $r=y$ (blue points) and $r=z$ (red points) axes for different laser intensities. The value of the diffusion coefficient expected from a fixed temperature (measured on average at $T=302.15$ K), viscosity ($\eta=0.8145 \times 10^{-3}$ N.s.m$^{-2}$) and particle diameter  ($d=940$ nm) is also displayed (black continuous line) with a light grey surface including errors in the determination of temperature ($\delta T = \pm 1$ K), viscosity ($\delta \eta = \pm0.018 \times 10^{-3} $ N.s.m$^{-2}$ due to $\delta T$) and particle size dispersion ($\delta d = \pm 50$ nm). The error bars on the experimental determinations of $D_y$ and $D_z$ are given at a 95\% confidence level from the linear regression taken on the variances of the displacement $\sigma^2(y(\Delta+t) - y(t))$ along $y$, and $\sigma^2(z(\Delta+t) - z(t))$ along $z$, respectively. These variances are shown on (b), noting $\sigma^2_r(\Delta) = \sigma^2(r(\Delta+t) - r(t))$ with $r=y$ (blue triangles) and $r=z$ (red squares). These variances, plotted as function of time lag $\Delta$, are calculated for 7 experiments done under the different laser intensities shown on panel (a).}
  \label{figS1}
\end{figure}

This variance can be calculated from the difference between the experimental time ensemble average MSD and the square of the mean displacement measured over $\Delta$ as:
\begin{equation}
\sigma^2(r(\Delta+t) - r(t)) = \langle \overline{\delta r_i^2(\Delta)} \rangle - \langle r(\Delta+t) - r(t) \rangle^2, \label{varMSD}
\end{equation}
where one recovers on $\langle \overline{\delta r_i^2(\Delta)} \rangle$ the ballistic contribution of the external force field. 

Fitting the variance respectively in the $y$ and $z$ directions with a chosen time lag $\Delta$ measures the diffusion coefficient $D_y$ and $D_z$. The results are shown in Fig. \ref{figS1} for different illumination powers, hence different strengths of radiation pressure exerted along the $y$ axis. The results clearly show, within error bars, that $D_y\sim D_z$ and that they fall in good agreement, within error bars (see caption), with the theoretical value of the diffusion coefficient $D=k_B T/ \gamma$ expected with a measured temperature $T=302.15$ K and a Stokes friction drag $\gamma$ evaluated from the corresponding viscosity of water and the known diameter of the colloidal sphere. This result is important since it provides a valid value (including uncertainties) for the Stokes friction drag, necessary in Eq. (\ref{shift}) for converting measured spatial displacements into force strength signals within well determined confidence intervals.  

\section{Appendix E: Ergodicity for free Brownian motion}

We here consider a one-dimensional free Brownian motion $x(t)$ described, in the overdamped limit, by the simple Langevin equation:
\begin{equation}
\dot x(t) = U(t)
\end{equation}
where $U(t)$ corresponds to a thermal noise with $\langle U(t) \rangle = 0$ and $\langle U(t_1)U(t_2) \rangle = \frac{2k_BT}{\gamma}\delta(t_1-t_2)$ where $k_{\rm B}$ is the Boltzmann constant, $T$ the temperature of the noise, and $\gamma$ the Stokes friction drag acting on the Brownian object considered. The ensemble average $\langle \cdots\rangle$ is done over all the realizations of the stochastic variable. Having:
\begin{equation}
\delta x(t) = x(t) - x(0) = \int_0^{t}dx(t)= \int_0^tU(t')dt' ,
\end{equation}
and setting $x(0)=0$, the mean-square displacement (MSD) is given by the Einstein equation:
\begin{equation}
\langle x(t)^2\rangle = \frac{2k_B T}{\gamma} t = 2Dt  \label{einstein}
\end{equation}
The position correlation function $C_x(t_1,t_2)$ can then be calculated:
\begin{equation}
C_x(t_1,t_2) = \langle x(t_1)x(t_2) \rangle
\end{equation}
If we suppose that $t_2 > t_1$:
\begin{equation}
\begin{aligned}
\langle x(t_1)x(t_2) \rangle &= \langle x(t_1) \cdot [x(t_1) + (x(t_2) - x(t_1))]\rangle \\
& = \langle x(t_1)^2 \rangle + \langle x(t_1) \cdot(x(t_2) - x(t_1)) \rangle\\
& = 2Dt_1 + \langle \int_0^{t_1} U(t)dt \int_{t_1}^{t_2} U(t)dt \rangle \\
& =  2Dt_1
\end{aligned}
\end{equation}
since the two integrals do not overlap.

Similarly, if $t_1 > t_2$, then $C_x(t_1,t_2) = 2Dt_2$. Both cases can be described in a unified manner with:
\begin{equation}
\begin{aligned}
C_x(t_1,t_2) & = \langle x(t_1)x(t_2) \rangle = 2D\min[t_1,t_2] \\
& = D(t_1 + t_2 - \left| t_1 - t_2 \right|).  \label{corrt1t2}
\end{aligned}
\end{equation}

We now calculate explicitly the ergodic parameter $\epsilon$ defined in Eq.(\ref{eqEB}) in the main text. First, we remind that $ \langle x(\Delta)^2\rangle = \langle \overline{\delta x^2 (\Delta)}\rangle$ where $\overline{\delta x^2 (\Delta)}$ is the time average MSD taken over the duration $\mathcal{T}$
\begin{equation}
\overline{\delta x^2 (\Delta)} = \frac{1}{\mathcal{T}-\Delta}\int_0^{\mathcal{T}-\Delta} (x_i(t'+\Delta) - x_i(t'))^2 dt'
\label{TAMSDx}
\end{equation}
and $\langle \overline{\delta x^2 (\Delta)}\rangle = 2D\Delta $ the corresponding ensemble average performed over all available $\overline{\delta x^2 (\Delta)}$. 

\begin{figure}[htb]
  \centering{
    \includegraphics[width=0.7\linewidth]{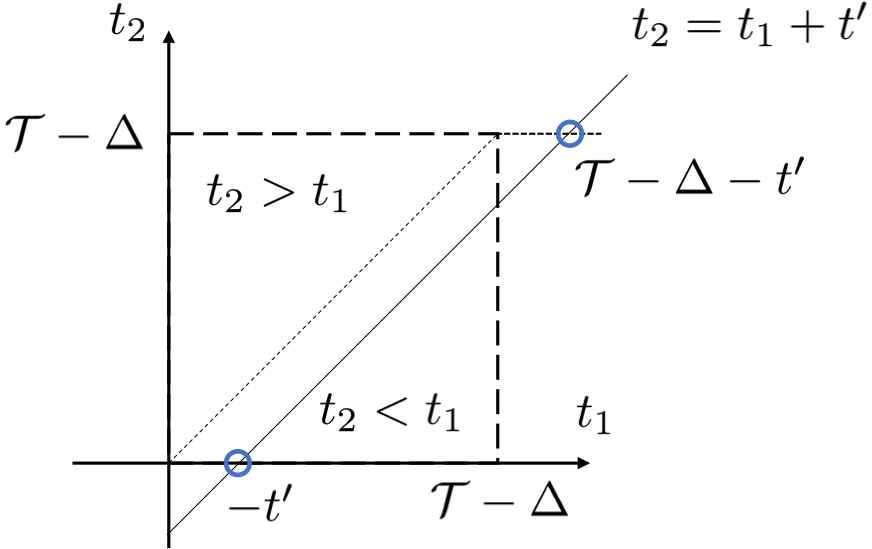}}
  \caption{Integration surface for Eq. (\ref{intalpha}) on which the two sectors $ [ t_2>t_1 ] $ and $ [ t_2<t_1 ]$ are distinguished. This defines the appropriate change of variables $(t_1,t_2)\leftrightarrow (t_1,t^\prime)$, with the line $t_2=t_1+t^\prime$ crossing the $t_2=0$ axis at $-t^\prime$ and the $t_2=\mathcal{T}-\Delta$ axis at $\mathcal{T}-\Delta+t^\prime$. }
  \label{figS2}
\end{figure}

The ergodic parameter $\epsilon$ is built on the variance of time average MSD $\sigma^2(\overline{\delta x^2(\Delta)}) = \langle \overline{\delta x^2(\Delta)}^2 \rangle - \langle \overline{\delta x^2(\Delta)} \rangle^2$. Looking at 
\begin{equation}
\begin{aligned}
\langle \overline{\delta x^2(\Delta)}^2 \rangle &= \frac{1}{(\mathcal{T}-\Delta)^2} \int_0^{\mathcal{T} - \Delta} dt_1 \int_0^{\mathcal{T} - \Delta} dt_2 \\
&\langle (x(t_1+\Delta) - x(t_1))^2(x(t_2 + \Delta) - x(t_2))^2 \rangle,
\end{aligned}
\end{equation}
we will use Wick's relation for normally distributed ensembles of displacements:
\begin{equation}
\begin{aligned}
&\langle x(t_1) x(t_2) x(t_3) x(t_4)\rangle = \langle x(t_1) x(t_2)\rangle \langle x(t_3) x(t_4) \rangle + \\
&~~~~\langle x(t_1) x(t_3)\rangle \langle x(t_2) x(t_4)\rangle + \langle x(t_1) x(t_4)\rangle\langle x(t_2) x(t_3)\rangle .
\end{aligned}
\end{equation}
Using Eq. (\ref{einstein}) and Eq. (\ref{corrt1t2}), one gets:
\begin{equation}
\begin{aligned}
&\langle (x(t_1+\Delta) - x(t_1))^2(x(t_2 + \Delta) - x(t_2))^2 \rangle   \\
&= \langle (x(t_1+\Delta) - x(t_1))^2\rangle \langle (x(t_2+\Delta) - x(t_2))^2\rangle\\
&+2\langle (x(t_1+\Delta) - x(t_1))(x(t_2+\Delta) - x(t_2))\rangle^2   \\
&  = 4D^2\Delta^2 + 2\cdot D^2 \cdot \alpha^2,
\end{aligned}
\end{equation}
with:
\begin{equation}
\begin{aligned}
\alpha =  |t_1-t_2 +\Delta | +|t_1-t_2 - \Delta | -2|t_1 - t_2 |
\end{aligned}
\end{equation}
The term $4D^2\Delta^2$ is cancelled in the variance by $\langle \overline{\delta x^2(\Delta)} \rangle^2$ and one is therefore left to calculate
\begin{equation}
\begin{aligned}
\sigma^2 \left(\overline{\delta x^2(\Delta)} \right) = \frac{2D^2}{(\mathcal{T}-\Delta)^2} \int_0^{\mathcal{T} - \Delta} dt_1 \int_0^{\mathcal{T} - \Delta} dt_2  \ \cdot \ \alpha^2.  \label{intalpha}
\end{aligned}
\end{equation}
To do so, we use the change of variables described in Fig. \ref{figS2} that splits the integration over the two distinct sectors $t_2<t_1$ and $t_2>t_1$. For the $t_2<t_1$ sector, we have
\begin{equation}
\begin{aligned}
\int_{-(\mathcal{T}-\Delta)}^{0}  dt^\prime \int_{-t^\prime}^{\mathcal{T} - \Delta} dt_1 (\Delta -t^\prime +|t^\prime +\Delta| +2t^\prime)^2 \label{Sec1}
\end{aligned}
\end{equation}

and for the $t_2>t_1$ sector
\begin{equation}
\begin{aligned}
\int_{0}^{\mathcal{T}-\Delta} dt^\prime \int_{0}^{\mathcal{T} - \Delta-t^\prime} dt_1
(|t^\prime-\Delta| + t^\prime +\Delta -2t^\prime)^2.  \label{Sec2}
\end{aligned}
\end{equation}
The absolute value integrands are finally evaluated using a last change of variable $v=t^\prime +\Delta$ on Eq.(\ref{Sec1}) and $w=t^\prime-\Delta$ on Eq.(\ref{Sec2}) that restricts the integration to $v>0$ and $w<0$ values, and eventually leads to: 
\begin{equation}
\begin{aligned}
& \int_0^{\mathcal{T} - \Delta} dt_1 \int_0^{\mathcal{T} - \Delta} dt_2  \ \cdot \ \alpha^2 =  \\
&\int_{0}^{\Delta} dv \cdot 4v^2\cdot (\mathcal{T}+v-2\Delta)+ \int_{-\Delta}^{0} dw \cdot 4w^2\cdot (\mathcal{T}-w-2\Delta).
\end{aligned}
\end{equation}
In the long-time limit $\mathcal{T}/\Delta \rightarrow \infty$:
\begin{equation}
\begin{aligned}
&\mathcal{T}+v-2\Delta \sim \mathcal{T}-w-2\Delta \sim \mathcal{T}  \\
&\frac{1}{(\mathcal{T}-\Delta)^2}  \sim \frac{1}{\mathcal{T}^2}, 
\end{aligned}
\end{equation}
so that Eq.(\ref{intalpha}) simplifies to:
\begin{equation}
\begin{aligned}
\frac{2D^2}{(\mathcal{T}-\Delta)^2} \int_0^{\mathcal{T} - \Delta} dt_1 \int_0^{\mathcal{T} - \Delta} dt_2  \ \cdot \ \alpha^2  \sim  \frac{16 D^2\Delta^3}{3 \mathcal{T}}.
\end{aligned}
\end{equation}

The ergodic parameter $\epsilon= \sigma^2 \left(\overline{\delta x^2(\Delta)} \right) / \langle \overline{\delta x^2(\Delta)}\rangle^2$ then simply writes in the long-time limit as:
\begin{equation}
\lim_{\mathcal{T}/\Delta \rightarrow \infty} \epsilon(\Delta) = \frac{\frac{16D^2}{3\mathcal{T}}\Delta^3}{4D^2\Delta^2} = \frac{4\Delta}{3\mathcal{T}}.
\end{equation}
This scaling is discussed already in \cite{Barkai2014} as a sufficient condition for ergodicity. As we demonstrate in the main text, this scaling is clearly verified in our experiments, proving the ergodic character of our colloidal system. 

\section{Appendix F: Correcting the ergodic parameter from tracking errors}

With such tracking-error corrections, we can give a relation between the experimentally estimated ergodicity breaking parameter $\epsilon (\Delta)_{exp}$ and the real one $\epsilon (\Delta)_0$, starting from the definition of the $EB$ parameter that involves the MSD according to:
\begin{equation}
\epsilon (\Delta)_{exp} = \frac{\langle \left( \overline{\delta y_i^2(\Delta)}_0 + \varepsilon_i \right)^2 \rangle}{\langle \overline{\delta y_i^2(\Delta)} \rangle_{exp}^2} - 1.  \label{ergocomp}
\end{equation}
Defining the ratio
\begin{equation}
\phi = \frac{\langle \overline{\delta y_i^2(\Delta)} \rangle_{0}}{\langle \overline{\delta y_i^2(\Delta)} \rangle_{exp}}  \label{phi}
\end{equation}
which, from Eq. (\ref{error}), can be written as a function of $\Delta$:
\begin{equation}
\phi(\Delta) = \frac{1}{1+b/2D\Delta}, \label{phi2}
\end{equation}
allowing us to rewrite Eq. (\ref{ergocomp}) as:
\begin{equation}
\epsilon (\Delta)_{exp} = \phi^2 \epsilon (\Delta)_0 + (\phi^2 - 1) + \frac{2\langle\varepsilon_i \overline{\delta y_i^2(\Delta)}_0 \rangle + \langle \varepsilon_i^2 \rangle}{\langle \overline{\delta y_i^2(\Delta)} \rangle_{0}^2} \phi^2.
\end{equation}

Assuming that $\varepsilon_i$ and $\overline{\delta y_i^2(\Delta)}_0$ are not correlated leads to $\langle \varepsilon_i \overline{\delta y_i^2(\Delta)}_0 \rangle = b \langle  \overline{\delta y_i^2(\Delta)}_0 \rangle$, so that:
\begin{eqnarray}
\epsilon (\Delta)_{exp} = \phi^2 \left( \epsilon (\Delta)_0 + \frac{\sigma^2(\varepsilon_i)}{(2D\Delta)^2} \right),  \label{EBfin}
\end{eqnarray}
considering that: 
\begin{equation}
\langle\varepsilon_i^2\rangle = \langle\varepsilon_i\rangle^2 + \sigma^2(\varepsilon_i) = b^2 +  \sigma^2(\varepsilon_i).
\end{equation}

This relation has an important consequence when discussing the time evolution of the ergodic parameter $\epsilon$. Indeed, as seen from Eq. (\ref{EBfin}), for sufficiently large time lags $\Delta \gg \Delta t$ with respect to the sampling time, the error term vanishes as $\phi \rightarrow 1$. In such conditions, $\epsilon (\Delta)_{exp} \sim \epsilon (\Delta)_0$. But when evaluating $\epsilon$ in the regime of small time lags with $\Delta \sim \Delta t$, the experimental result $\epsilon (\Delta)_{exp}$ turns out to be larger than $\epsilon (\Delta)_0$ precisely because of the influence of the tracking errors. What can be taken for the signature of some breaking of ergodicity at small time lags is simply, in this case, an artefact of the experiment which is simply accounted for using Eq. (\ref{EBfin}). 

In order to evaluate Eq. (\ref{EBfin}) as presented in Fig. \ref{fig2} (b) in the main text, we estimate $\phi$ using Eq. (\ref{phi2}) and evaluate $\sigma^2(\varepsilon_i)$ in two steps. First, we assume that $\overline{\delta y_i^2(\Delta)}_{exp}$ and $\overline{\delta y_i^2(\Delta)}_{0}$ are uncorrelated, so that according to Eq. (\ref{deltaExp}):
\begin{eqnarray}
\sigma^2(\varepsilon_i) = \sigma^2 \left(\overline{\delta y_i^2(\Delta)}_{exp} \right) + \sigma^2 \left(\overline{\delta y_i^2(\Delta)}_{0}\right).
\end{eqnarray}
We then take Eq. (\ref{eqEB}) in the smallest $\Delta \sim \Delta t$ limit, for which $\sigma^2 \left( \overline{\delta y_i^2 (\Delta t)_0}  \right)\sim\langle \overline{\delta y_i ^2(\Delta t)_0} \rangle^2  \times 4\Delta t / 3\mathcal{T} \sim (\Delta t)^2/\mathcal{T} \sim 0$ considering that $\langle \overline{\delta y_i ^2(\Delta)_0} \rangle^2=2D\Delta t$. This leads eventually to:
\begin{eqnarray}
\sigma^2(\varepsilon_i) \sim \sigma^2 \left(\overline{\delta y_i^2(\Delta t)}_{exp} \right).
\end{eqnarray}

\begin{figure}[htb]
  \centering{
    \includegraphics[width=0.6\linewidth]{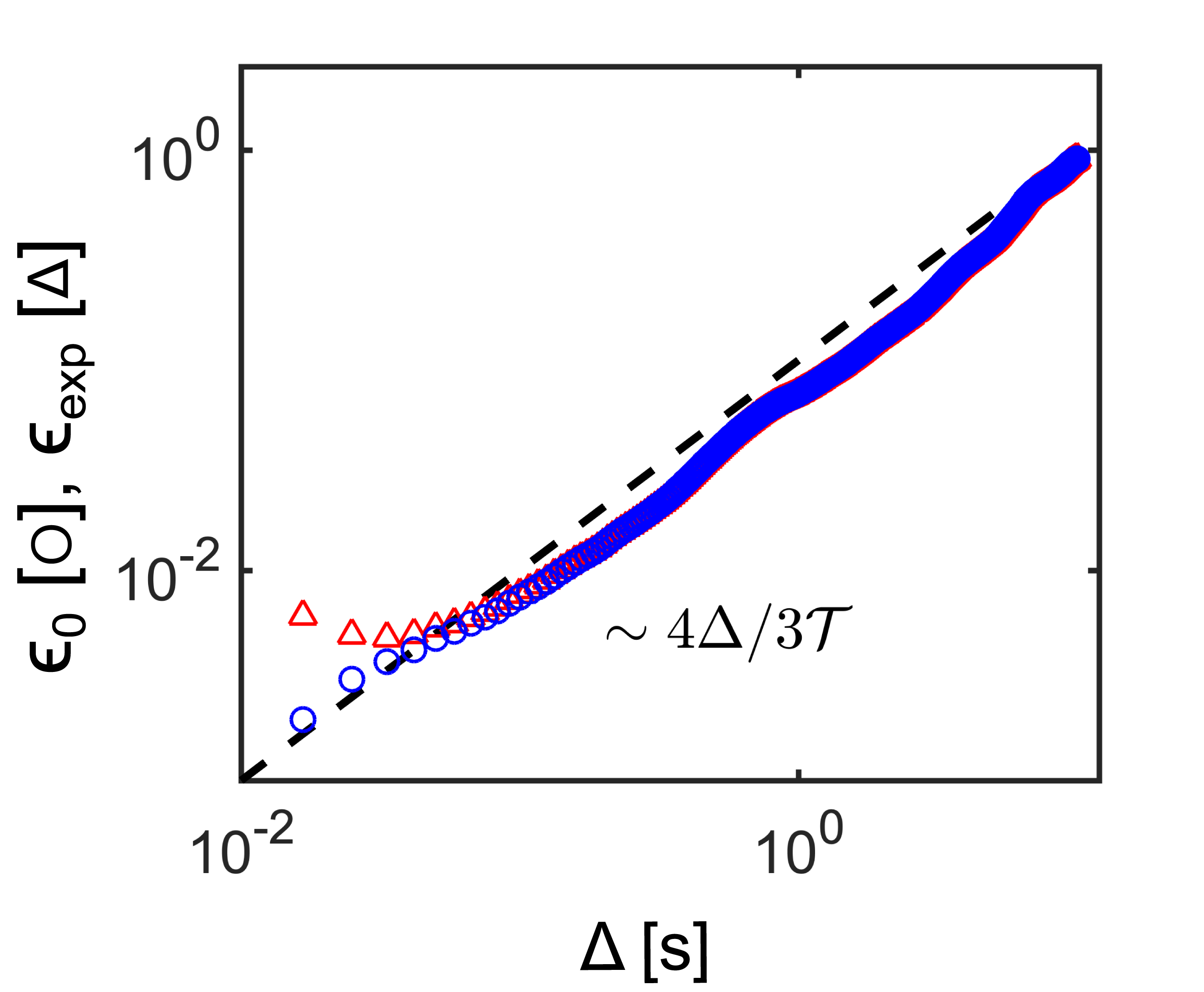}}
  \caption{Comparison between ergodic parameters $\epsilon$ before ($\epsilon_{exp} (\Delta)$, red triangles) and after correction ($\epsilon_0 (\Delta)$, blue circles), revealing a clear difference at small time lags $\Delta$ but almost none for larger $\Delta$. The superimposed dashed line corresponds to the ergodic long-time limit of free Brownian motion calculated in Appendix E. These data correspond to the lowest intensity ($0.006$ kW/cm$^{2}$) experiment shown in Fig. \ref{fig2}(c).}
  \label{figS3}
\end{figure}

\section{Appendix G: Radiation pressure force uncertainty}

The global uncertainty related to a radiation pressure force measurement $F_0$ involves different sources of errors: $(i)$ of systematic nature regarding the determination of the colloidal particle size, the temperature (hence viscosity) of water and the laser beam waist $w_0$ and center $z_0$, and $(ii)$ of statistical nature related to the standard error associated with the ensemble averaging of the displacement distribution. These different errors combine in a global uncertainty as:
\begin{equation}
\delta  F_0 = \sum_u (\delta F_0)_u = \sum_u \left| \frac{\partial F_0}{\partial u} \right| \delta u.
\end{equation}

From Eq. (\ref{estim}), we know that the $F_0$ can be written as:
\begin{equation}
F_0 = \frac{1}{\alpha}\cdot \frac{\gamma}{\Delta t}\langle\Delta y_n \rangle,
\end{equation}
where $\alpha = \langle \exp \left( -2(z_n - z_0)^2/w_0^2 \right) \rangle$. The error on $\alpha$ comes from the determination error of $z_0$ and $w_0$, which yields a relative error of $\varphi_{\alpha} = \delta \alpha / \alpha \sim 0.4\%$. Within the Stokes drag term $\gamma = 6\pi\eta a$, both systematic errors are at play. The $\pm 1$ K error on temperature around a mean value of $302.15$ K leads to a relative error in the viscosity $\varphi_{\eta} \sim 2.4\%$, and the $50$ nm standard deviation in particle size given by the manufacturer corresponds to a relative error in particle diameter of $\varphi_a \sim 5.3\%$. The standard error in the displacement distribution is calculated from the ensemble of $\{ \Delta y_n \}$, so that, using the maximum error estimation method, the uncertainty writes as follows:
\begin{eqnarray}
\delta F_0 &=& (\delta F_0)_{\alpha} + (\delta F_0)_{\eta} + (\delta F_0)_{a} + (\delta F_0)_{\langle\Delta y_n \rangle}  \nonumber \\
&=& F_0 \cdot (\varphi_{\alpha} + \varphi_{\eta} + \varphi_a) + \frac{\gamma}{\alpha\Delta t} \frac{\sigma(\Delta y_n)}{\sqrt{N}},
\end{eqnarray}
where $\sigma(\Delta y_n)$ is the sample standard deviation defined as:
\begin{equation}
\sigma(\Delta y_n) = \sqrt{\frac{\sum_n (\Delta y_n - \langle \Delta y_n \rangle)^2}{N - 1} }.
\end{equation}

\section{Appendix H: Mie calculation uncertainty}

Following \cite{CanaguierPRA2014}, radiation pressure forces are calculated in the Mie regime as $F_{pr} = (n_w I/c)\cdot Q_{pr}$, where $n_w$ is the refractive index of water, $I$ is the intensity of the illuminating laser, $c$ the velocity of light and $Q_{pr}$ is the calculated pressure cross section. Here, the uncertainty comes from the laser intensity because of inevitable errors made in measuring $(i)$ the laser intensity, and $(ii)$ the optical waist of the laser beam inside the cuvette. With $I = 2P/\pi w_0^2$, the laser intensity uncertainty is therefore given by:
\begin{equation}
\begin{aligned}
\delta I &= \frac{2}{\pi w_0^2}\delta P + 2 \frac{2P}{\pi w_0^3} \delta w_0 = \frac{2P}{\pi w_0^2} \frac{\delta P}{P} + \frac{2P}{\pi w_0^2} \cdot 2\frac{\delta w_0}{w_0} \\
&= I \left( \frac{\delta P}{P} +  2\frac{\delta w_0}{w_0} \right).
\end{aligned}
\end{equation}

In our experiments, we have a relative error in the power determination of ca. $5\%$ and a relative error in the waist measurement of ca. $4.9\%$ through the Gaussian fitting of the intensity profile with $95\%$ confidence level (see main text).

\bibliography{biblio-ergodic}

\end{document}